\documentclass[twocolumn,showpacs,preprintnumbers,pre,floatfix]{revtex4}
\usepackage{graphicx}
\usepackage{color}
\usepackage{epsfig}
\usepackage{amsmath}
\usepackage{amssymb}
\usepackage{verbatim}
\usepackage{graphics}
\usepackage{natbib}

\usepackage{float}
\floatstyle{ruled}

\usepackage{pstricks}
\usepackage{eepic}

\begin{document}

\title{Stress Response in Confined Arrays of Frictional and Frictionless
  Particles}

\author{Abdullah Cakir}
\author{Leonardo E. Silbert}

\affiliation{Department of Physics, Southern Illinois University, Carbondale,
  Illinois 62901, USA}

\begin{abstract}
  Stress transmission inside three dimensional granular packings is
  investigated using computer simulations. Localized force perturbation
  techniques are implemented for frictionless and frictional shallow, ordered,
  granular arrays confined by solid boundaries for a range of system
  sizes. Stress response profiles for frictional packings agree well with the
  predictions for the semi-infinite half plane of classical isotropic
  elasticity theory down to boxes of linear dimensions of approximately forty
  particle diameters and over several orders of magnitude in the applied
  force. The response profiles for frictionless packings exhibit a
  transitional regime to strongly anisotropic features with increasing box
  size. The differences between the nature of the stress response are shown to
  be characterized by very different particle displacement fields.
\end{abstract}
\maketitle

\section{Introduction}
The manner in which granular materials respond to external loads belongs to a
class of problems related to mechanical stability. Questions surrounding
mechanical stability of particulate media have broad economical,
environmental, and societal applications. These include the common practical
problem of a collapsing grain silo, numerous natural events such as avalanches
and mud slides, along with other unexpected destructive occurrences like dam
breakage. Many such events are reported annually and readily convey the
dramatic, and often catastrophic, effects that are primarily due to the
failure of various granular systems.

The general physical theme that connects such a disparate range of problems is
that of determining the underlying stress state of the system. Given
information on the distribution of stresses within the system, can we predict
whether such systems will remain mechanically stable, and what are the
conditions that might promote or inhibit failure? One possibility is to employ
established theories such as continuum, linear, elasticity theory
\cite{Johnson}. Although elasticity theory applies over a wide range of
situations, deviations from classical elasticity theory show up on a regular
basis as some of the assumptions that underpin the theory break down. These
problems have largely been emphasized over the past 15 years or so in studies
of packings of granular materials. The essential point to note is that
granularity, or the discreteness of the particles that constitute the packing,
play a vital role in the resulting stress characteristics of the system due to
a lack of a separation of length scales. Thus, it is becoming increasingly
necessary to probe the relevant limits of classical elasticity theory and its
range of valid application. For instance, classical elasticity is often
assumed to remain valid not only at macroscopic scales to infer stress
properties from displacement fields during loading events of the lithosphere
\cite{10.1111/j.1365-246X.2003.02150.x}, but also at the microscopic level of
interrogation in studies of nanoscopic indentation in relation to physical
metallurgy
\cite{10.1134/S1063785008070201,springerlink:10.1007/s11012-006-9018-6}.

A convenient method to test the range of validity of classical elasticity
theory in granular systems that has been utilized over the past decade is the
localized force perturbation technique. This procedure has the particular
benefit that for an isotropic elastic material the stress profile in response
to localized forcing - Green function response - can be determined without the
need for any free parameters. We review this result for the case of three
dimensions - Boussinesq equations - in the following section (and see the
Appendix). To summarize efforts to date, studies on two- and three-dimensional
granular packings indicate that disordered and frictional systems tend to
follow the predictions of isotropic elasticity theory. Whereas, ordered
packings composed of frictionless particles result in strongly anisotropic
stress profiles that are not consistent with a linear, isotropic, elastic
model. Such anisotropic profiles (see below for a more qualitative definition)
have been debated in relation to anisotropic elasticity theory based on
differential equations of the elliptic class \cite{PhysRevE.67.031302}, and
alternative descriptions of the hyperbolic wave variety
\cite{10.1051/jp1:1995157,10.1051/jp1:1997126}. To distinguish between the
traditional isotropic result and anisotropic stress profiles we refer to these
as \emph{one-peak} and \emph{two-peak} response profiles respectively, in
reference to two dimensional experiments and simulations that have reported
these observations
\cite{10.1007/s100350100086,PhysRevLett.87.035506,Nature,1742-5468-2006-09-P09003,PhysRevE.77.041303,springerlink:10.1007/s10035-010-0215-6,PhysRevE.83.021304}.

In recent computer simulations the role of structure for frictionless granular
packings on the stress response in a pseudo-infinite medium was investigated
\cite{10.1007/s10035-009-0156-0}. These studies showed that for ordered and
quasi-ordered arrangements of frictionless grains, the response function was
two-peak but consistent with the framework of an anisotropic elastic theory. A
phenomenological model was employed to fit these anisotropic profiles that
necessarily captured the two-peak character of the response profiles. As the
structural disorder of the packing was increased, the stress response crossed
over to a one-peak response, more in line with isotropic behavior.

The problem addressed here is to determine how confinement affects the stress
response of granular packings. In most previous studies samples were used of
sufficient size that the perturbed packings could be considered as
semi-infinite in extent. In some cases the roughness of the supporting base in
packings prepared under gravity was seen to influence the magnitude of the
stress response, although the response profiles retained isotropic, elastic
character \cite{10.1007/s101890170019}. We also highlight two particularly
relevant efforts that have been proposed to describe stress response specific
to two dimensional systems. Extensions to three dimensions and especially for
fully confined, shallow packings remains an ongoing effort. A non-linear
elasticity formalism was developed \cite{PhysRevE.77.031303} that captures the
main features of experimental results for two dimensional granular packings
\cite{PhysRevLett.87.035506}. The isotropic result is recovered through a
multipole expansion of the normal stress response component $\sigma_{zz}$ of
classical elasticity. This theory was also able to describe anisotropic
profiles by including generalized nonlinear strain components through the
introduction of phenomenologically defined elastic constants. A future
extension of this particular formalism may be useful in describing the stress
response for a wide range of granular systems in both two and three
dimensions. Additionally, force perturbation simulations of confined
Lennard-Jones glasses \cite{PhysRevB.70.014203} found a range of stress
response profiles that appear, at least superficially, similar to some of the
isotropic results we present here. Moreover, in some of the above examples a
stress response crossover was reported either with system size and/or packing
arrangement. This suggests that stress response investigations on granular
packings can provide a route to understanding stress states in a range of
systems where the discreteness of the constituent particles come into play.

However, there currently remains a distinct lack of a systematic study on the
role that system size plays in modifying the stress response in granular
materials. We propose that such a study is needed with the ever-growing
emphasis on the design of small devices where the particulate nature of the
material eventually dominate the system properties. Moreover, from this study
we can estimate relevant length scales over which continuum elasticity theory
remains valid, and below which alternative descriptions may be needed. Thus,
our goal here is to provide a qualitative guide map to the typical system
parameters for which classical elasticity theory is likely to apply and those
for which more advanced techniques may be needed to predict their properties.
 
This paper is organized as follows. In the Section
\ref{sec:BoussinesqEquations} we summarize the Boussinesq equations of
classical, isotropic, linear elasticity theory for an infinite half-plane, and
discuss other possibilities that predict anisotropic stress response
profiles. In Section \ref{simulations} we provide an overview of the
computational technique and the packing preparation protocol implemented in
this work. Section \ref{results} presents our main findings for the stress
response behavior in frictional and frictionless particle packings. We end
with conclusions and discussion.

\section{Boussinesq Equations}
\label{sec:BoussinesqEquations}
The equations of classical elasticity theory are derived from differential
equations of the elliptic class that rely on the identification of well
defined displacement fields of elements constituting the continuous elastic
medium \cite{Landau}. The analytical solutions for the problem of a localized
point force applied vertically at the surface of an elastic, isotropic,
infinite half-space are called the Boussinesq equations
\cite{Johnson,Landau}. In Cartesian coordinates, where the origin of the
coordinate system coincides with the application point of the vertical force,
$F_{\rm app}$, the components of the displacement fields are given by,
\begin{equation}
  \begin{tabular}{ccl}
    $u_{x}$ & $=$ & $\frac{F_{\rm app}}{4\pi G}\left[ \frac{xz}{\rho^{3}} -
      (1-2\nu)\frac{x}{\rho(\rho+z)}\right]$ \\ \\
    $u_{y}$ & $=$ & $\frac{F_{\rm app}}{4\pi G}\left[\frac{yz}{\rho^{3}} -
      (1-2\nu)\frac{y}{\rho(\rho+z)}\right]$ \\ \\
    $u_{z}$ & $=$ & $\frac{F_{\rm app}}{4\pi G}\left[\frac{z^{2}}{\rho^{3}} + 
      \frac{2(1-\nu)}{\rho}\right]$
  \end{tabular}
  \label{eq1}
\end{equation}
where $\rho = \sqrt{x^2+y^2+z^2}$. Both $G$, the modulus of rigidity, and
$\nu$, the Poisson ratio, are material parameters characterizing the bulk
properties of the elastic medium. Based on thermodynamic grounds
\cite{Landau}, the Poisson ratio is constrained by, $-1<\nu<0.5$, in three
dimensions. Through implementation of Hooke's law between stress and strain
components, one then obtains expressions for components of the stress
tensor. For the specific case when the force is applied in the `downwards' or
negative $z$-direction (see Fig.~\ref{fig1}) the normal stress in the
direction of the applied force, $\sigma_{zz}$, characterizes the response of
the system. We focus primarily on this component of the stress although we
also provide information on the normal stress components not in the direction
of the applied force, $\sigma_{xx}$ and $\sigma_{yy}$. We give expressions for
all three normal stress components
\begin{equation}
  \begin{tabular}{ccl}
    $\sigma_{xx}$&$=$&$\frac{F_{\rm app}}{2\pi}[\frac{1-2\nu}{r^2}((1-\frac{z}{\rho})\frac{x^2-y^2}{r^2}+\frac{z y^2}{\rho^{3}})-\frac{3zx^{2}}{\rho^{5}}]$ \\ \\
    $\sigma_{yy}$&$=$&$\frac{F_{\rm app}}{2\pi}[\frac{1-2\nu}{r^2}((1-\frac{z}{\rho})\frac{y^2-x^2}{r^2}+\frac{z x^2}{\rho^{3}})-\frac{3zy^{2}}{\rho^{5}}]$ \\ \\
    $\sigma_{zz}$&$=$&$\frac{3F_{\rm app}}{2\pi}\frac{z^3}{\rho^{5}}$ 
    \label{eq2}
  \end{tabular}
\end{equation}
where $r=\sqrt{x^2+y^2}$. 

The stress response scales linearly with the applied force. Deviations from
this indicate non-linear behavior, which, in the case of a particulate medium,
corresponds to non-affine, or irreversible, particle displacements in response
to the applied load. We reiterate that that there are no fitting parameters in
$\sigma_{zz}$ for the case of a semi-infinite half space. Therefore, we can
quantify the influence of boundaries, or confinement, through a comparison of
the Boussinesq result of Eq.~\ref{eq2}, with our computed stress response for
confined systems. Furthermore, we can also quantify deviations from isotropy
from the stress profiles. We find that although the application of Boussinesq
theory might at first seem to be inappropriate for the systems investigated
here, we find that for packings of sufficient linear size the classical theory
captures the essential features of the stress properties and we classify the
conditions for which isotropic elasticity theory are suitable.

\section{Simulation Methods}
\label{simulations}
\subsection{The Contact  Forces and Equations of Motion}
We implement a granular dynamics (GD) variant on molecular dynamics
simulations \cite{Allen:1989:CSL:76990} that has been designed to simulate
granular systems \cite{Plimpton:1995:FPA:201627.201628}.  Here we provide a
summary of the technique \cite{SilbertBook}. For this particular study we
focus on noncohesive, monodisperse sphere packings composed of elastic spheres
of diameter $d$ and mass $m$, that interact only on contact, through forces
that act in directions that are normal ${\bf n}$, and tangential ${\bf t}$, to
the contact plane, defined via
\begin{equation}
  \begin{tabular}{ccc}
    ${\bf{n}}$ & = & $\frac{{{\bf r}_{ij}}}{r_{ij}}$\\ \\
    ${\bf{t}}$ & = & ${\bf{1}}-{\bf{n}}$
    \label{eq3}
  \end{tabular}
\end{equation}
where ${\bf r}_{ij} = {\bf r}_{i} - {\bf r}_{j}$, is the separation between
particles $i$ and $j$, located at positions ${\bf r}_{i}$ and ${\bf r}_{j}$
respectively, and $r_{ij}=|{\bf r}_{ij}|$. ${\bf 1}$ represents the unit
vector.

The normal and tangential contact forces $F_{n,t}$ are based on a linear
spring-dashpot model characterized by normal and tangential linear spring
constants $k_{n,t}$, and damping factors $\gamma_{n,t}$, that account for
elastic deformation of the contact point and inelasticity respectively,
\begin{equation}
  \begin{tabular}{ccc}
    $F_{n}$ & = & $-k_{n}\delta_{ij} - m_{\rm eff} \gamma_{n}v^{n}_{ij}$\\ \\
    $F_{t}$ & = & $-k_{t}s_{ij}- m_{\rm eff} \gamma_{t}v^{t}_{ij}$.
    \label{eq4}
  \end{tabular}
\end{equation}
Here, $\delta_{ij} \equiv (r_{ij} - d)$ is the surface compression of two
particles undergoing a collision and $v^{n,t}_{ij}$ the relative
normal/tangential velocity. The effective mass, $m_{\rm eff} = m/2$, for the
systems studied here. The quantity $s_{ij}$ represents the integrated
displacement of the contact point while two particles remain in
contact. Friction is implemented through a local Coulomb yield criterion:
$F_{t} \leq \mu F_{n}$, for a given friction coefficient $\mu$. In this work,
$\mu$ takes on the values zero and one. Thence, the total contact force is
given by ${\bf{F}} = F_{n}{\bf{n}} + F_{t}{\bf{t}}$. Interactions between the
particles and the walls are given by similar expressions where the wall is
characterized by an effective mass of unity and the wall friction coefficient
$\mu_{W}$ takes on the same values as that of the particles. The results are
presented in simulation units. Length scales are given in units of $d$,
timescales in units of $\sqrt{d/g}$, where $g$ is the acceleration due to
gravity, forces in units of $mg$, stresses in units of $mg/d^{2}$, and
energies are given in units of $mgd$. Table~\ref{simulpar} shows the
simulation parameters used in this study. We point out that the values of the
spring constants are fixed throughout this study resulting in a particle scale
Poisson ratio of zero.
\begin{table}
  \caption
  {  \label{simulpar}
    Physical parameters used in this simulation study. $k_{n}$ and $k_{t}$ are
    normal and tangential spring constants in units of $mg/d$. $\gamma_{n}$ and
    $\gamma_{t}$ are normal and tangential damping factors in units of
    $\sqrt{g/d}$. For values used here the particle Poisson ratio is zero. The
    particle friction coefficient is $\mu$ and the
    coefficient of inelasticity $e$.
  }
\begin{tabular}{| l | l | l |}
\hline
Number of Particles & N & 1000 - 100000\\
\hline
Normal Stiffness & $k_{n}$ & 1 $\times 10^{5} mg/d$\\
\hline
Tangential Stiffness & $k_{t}$ & 1 $\times 10^{5} mg/d$\\
\hline
Normal Damping Coefficient & $\gamma_{n}$ & $50 \sqrt{g/d}$\\
\hline
Tangential Damping Coefficient & $\gamma_{n}$ & $\gamma_{n}/2$ \\
\hline
Normal Restitution Coefficient & $e$ & 0.88 \\
\hline
Particle Friction Coefficient & $\mu$ & 0, 1.0\\
\hline
Wall Friction Coefficient & $\mu_{W}$ & 0, 1.0\\
\hline
\end{tabular}
\end{table}

\subsection{Coarse Graining Procedure}

Stresses are computed from the contact forces. To enable an accurate
determination of stresses from microscopic to macroscopic length scales, we
follow the protocol developed by Goldhirsch and colleagues
\cite{10.1007/s10035-010-0181-z}. We implement a coarse graining procedure
into the microscopic stress expression \cite{PhysRevLett.89.084302}. The
coarse graining function chosen is a 3D Gaussian, and the resulting expression
for the stress components at some location ${\bf r}$ is,
\begin{equation}
  \sigma_{\alpha \beta}({\bf{r}})=\frac{1}{2}\sum_{i \neq j} F_{ij \alpha}
  r_{ij \beta} \int_{0}^{1} ds \frac{1}{ \pi^{3/2} \omega^{3}} e^{-(\mid
    {\bf r}-{{\bf r}_{i}} + s {{\bf r}_{ij}} \mid)^{2} /\omega^{2}},
  \label{eq5} 
\end{equation} 
where $\omega$ is the coarse graining length scale. ${\bf F}_{ij}$ represents
the force experienced by particle $i$ in contact with particle $j$, obtained
from Eq.~\ref{eq4}. The numerical prefactor $(\sqrt{\pi} \omega)^{-3}$ is
the normalization factor.

To implement Eq.~\ref{eq5} into our calculations we adopted a spatial
grid over which to evaluate the coarse grained stress. To achieve sufficient
spatial resolution we fixed the grid size $= 0.1d$ and $\omega = 1.0d$. (We
computed stresses for a range of $\omega$ and grid sizes for our 3D systems
and found that $0.5d < \omega < 5d$ provides a suitable plateau window
\cite{PhysRevLett.96.168001}, at an acceptable computational cost.)

\subsection{Packing Generation Protocol}

\begin{figure}[ht]
  \includegraphics[height=7cm,width=7cm]{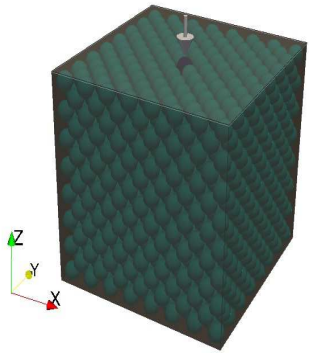}
  \caption{A computer simulation snapshot of a typical system studied in this
    work. The packing is composed of monodisperse spheres arranged into a face
    centered cubic array, confined by 4 side walls, a solid base with an open
    top. The application of a localized perturbative force is represented by
    the arrow and the coordinate system is defined in the figure. $x=0$ and
    $y=0$ define the center of box in the bottom of box. Gravity points in the
    $-z$ direction.}
  \label{fig1}
\end{figure}

In this work we investigate how confinement influences the nature of stress
response in granular packings. Because of the potentially large parameter
space we fixed many features of the packings and focus our attention on a subset
of variables. More specifically, we studied face centered cubic (FCC) ordered
arrays of monodisperse spheres confined within a walled box. To prepare the
initial configuration, we placed non overlapping particles at the lattice points
of an FCC structure at the hard sphere packing fraction, $\phi_{\rm FCC} =
\pi/\sqrt{18}$. The packings are confined within open-top boxes, consisting of
flat side walls and a bottom base.  We defined a coordinate system as shown in
Fig.~\ref{fig1}. The bottom plane of the box defines the $xy$-plane, while
gravity points in the $-z$ direction. The size of the simulation boxes were
chosen to be commensurate with the crystal structure. For the FCC structures
the lattice unit is defined as $a=\sqrt{2}d$. Most of the results presented
here are for boxes with square bases of side lengths, $L$, ranging from, $7a
\approx 10d$ to $70a \approx 100d$. We also studied rectangular shaped
boxes. The total number of particles $N$, was chosen by fixing the height,
$H$, of all packings to be, $H \approx 10d$.

After the particles were positioned at the lattice sites, we switched on
gravity and allowed the particles to settle until the energy per particle,
$E/N \approx 10^{-16}mgd$, reached numerical precision. This then defines our
initial configuration. We recorded the positions of the particles, the contact
forces between interacting particles, and consequently the stress,
$\sigma_{\rm i}$, of this initial state. We then identified the top middle
region of the packing, locating the particles in this region.

The second part of the simulation involves applying a downwards-pointing
localized perturbation force, $F_{\rm app}$, to the top middle particle(s),
and allowing the system to relax back to a mechanically stable state. We
varied the magnitude of this applied force over several orders of magnitude,
$0.01 mg \leq F_{\rm app} \leq 100 mg$, to check the limits of the linear
regime. Again, we tracked particle positions and contact forces, and the final
stress state $\sigma_{\rm f}$ \cite{normal}. The response function is
calculated as the difference between the final and initial stress states,
which we denote simply as,
\begin{equation}
  \sigma \equiv \sigma_{\rm f} - \sigma_{\rm i} .
  \label{eq6}
\end{equation}

To make a clearer connection with previous experimental studies and
visualizations \cite{PhysRevLett.87.035506,PhysRevLett.86.3308}.  we present
our results in a convenient format that can be readily compared with previous
studies and theoretical predictions. In particular, we make use of three
dimensional contour stress maps and two dimensional stress profiles which view
the stress response in a given plane of the box or angularly-averaged slice
across this plane, respectively. Given the box geometry that we employ, it is
convenient to visualize $\sigma_{\rm zz}$ as a stress map in the horizontal
plane. Using these plots it is relatively straightforward to distinguish
between isotropic and anisotropic response profiles. These typically show up
more clearly in stress profiles as single- and double-peak response profiles
respectively. In all our stress maps lighter shading indicates larger
stresses. We also present results on particle displacement fields taken in the
plane.

\section{Results}
\label{results}
\subsection{Frictional Packings}
\begin{figure}[htbp]
  \includegraphics[width=4cm]{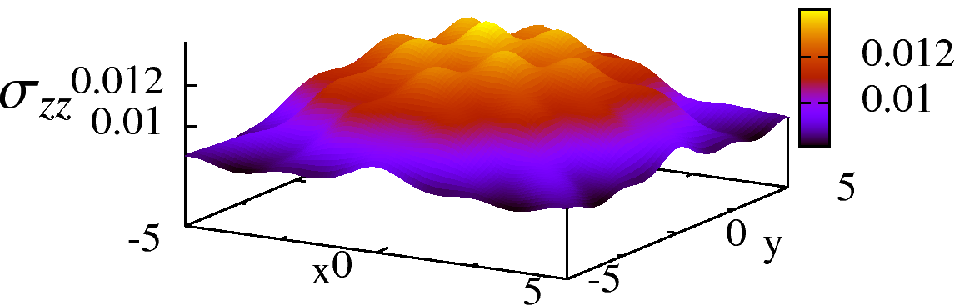}
  \includegraphics[width=4cm]{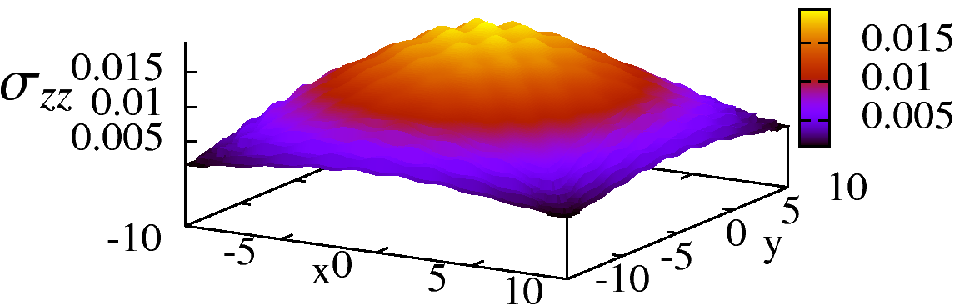}
  \includegraphics[width=4cm]{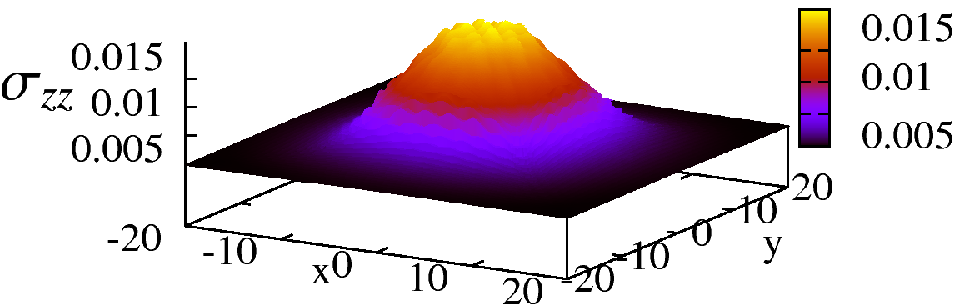}
  \includegraphics[width=4cm]{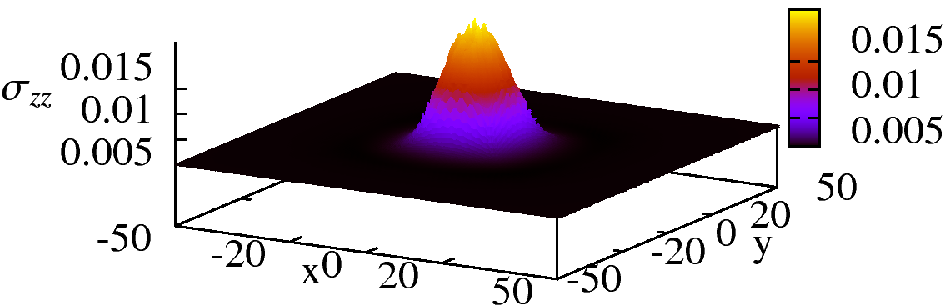}
  \caption{Frictional systems $\mu=1$. Stress response maps of $\sigma_{zz}$
    at the bottom boundary due to an applied force $F_{app} = 1mg$, for
    different system sizes, $10d \leq L \leq 100d$ indicated by scale in
    panels.}
  \label{fig2}
\end{figure}

Our first results show that ordered arrays of \emph{frictional} granular
packings display isotropic elastic-like stress response behavior as seen in
Fig.~\ref{fig2}. These stress maps convey the ``single-peak'' nature
of the stress response measured at the bottom of the packing for different
system sizes for one value of the applied force, $F_{app} = 1mg$. Thus, we
immediately note that the Boussinesq equation appears to describe confined
systems. However, we find that system size and force magnitude influence this
picture. We quantify the regime over which the classical expression remains
suitably valid in two ways: Firstly, we compare our stress profiles to the
Boussinesq expression and extract the full width at half maximum. Secondly, we
identify the linear regime as a function of the applied force.

\begin{figure}[htbp]
(a) \hfil
  \includegraphics[width=3.5cm]{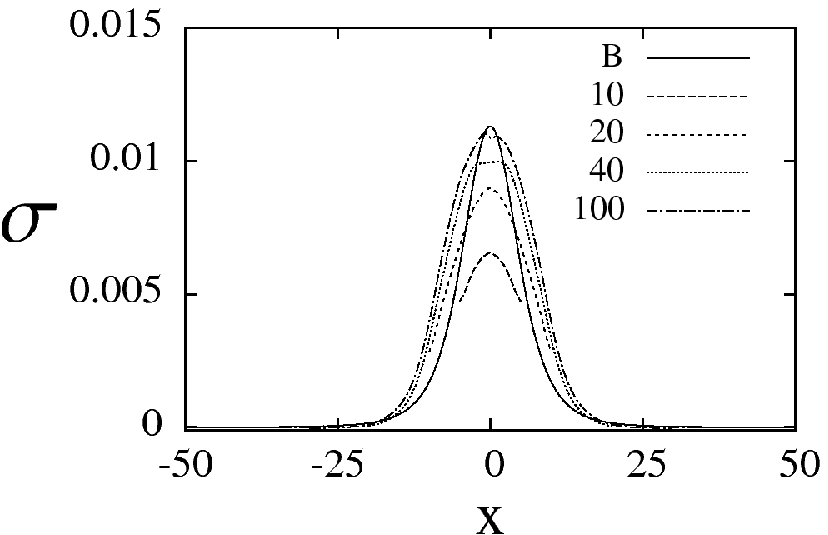}
(b) \hfil
  \includegraphics[width=3.5cm]{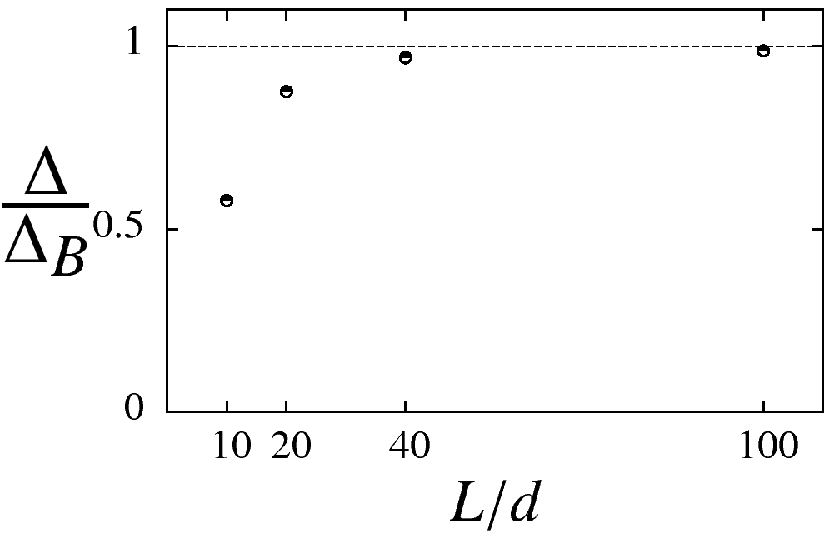}
  \caption{(a) Averaged stress response profiles, $\sigma$, for one value of
    the applied force $F_{app} = 1mg$ for systems of fixed height $H\approx
    10d$ and varying side lengths $L = 10d$ (long dash line), $20d$ (thin
    solid line), $40d$ (dashed), $100d$ (dotted), and B (solid line), the
    Boussinesq result. (b) $\Delta$ full width at half height of the stress
    profiles for the simulation data (symbols) relative to the theoretical
    result $\Delta_{B}$. Perfect agreement indicated by dashed line.}
  \label{fig3}
\end{figure}

To cast this data into a more familiar form we circularly average the stress
map data of Fig.~\ref{fig2} to construct stress profiles and directly
compare with the predictions of Eq.~\ref{eq2}, from which we also obtain the
full width of the response peaks. In Fig.~\ref{fig3}(a) we show our averaged
response profiles for systems of different height, clearly indicating a
single-peak response consistent with the classical theory shown by the thick
solid line.
We note that for the smallest system ($L=10d$) the stress profiles do not
accurately conform to the Boussinesq form. To better quantify deviations from
Eq.~\ref{eq2} we extract the full width at half height, $\Delta$, and compare
with the theoretical result $\Delta_{B} = 1.113 H$, shown in
Fig.~\ref{fig3}(b). We find that for $L > 40d$, the single-peak response
profiles are suitably described by Eq.~\ref{eq2}, with better agreement for
larger system size as expected.

\begin{figure}[htbp]
  \includegraphics[width=7cm]{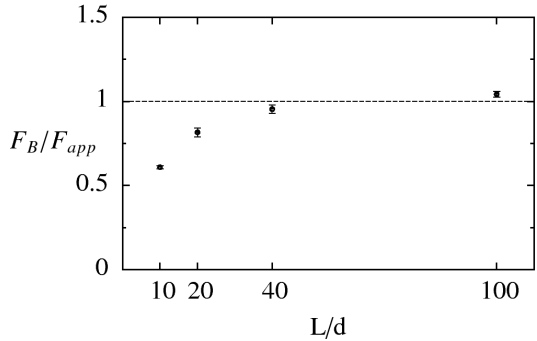}
  \caption{Comparison between the actual applied force $F_{app}=1mg$ versus
    $F_{B}$, the fitted force using the Boussinesq equation, Eq.~\ref{eq2}, to
    fit the stress profile data at the bottom of a frictional granular packing
    for boxes of different sizes, with side lengths $10d \leq L \leq
    100d$. Perfect agreement indicated by dashed line.}
  \label{fig4}
\end{figure}

Thus, we have identified a minimum system size for which the classical
expression describes the data accurately, $L \gtrsim 40d$. To further test our
results with Eq.~\ref{eq2}, we again fit the data and now allow the applied
force parameter in Eq.~\ref{eq2} to vary as a fitting parameter, denoted here
as $F_{B}$, which we then compared to the actual force applied during the
simulation, $F_{app}$. Our results shown in Fig.~\ref{fig4} clearly indicate
that confined systems of box sizes greater than $L \gtrsim 40d$ conform to the
semi-infinite system size result.

\begin{figure}[htbp]
  \includegraphics[width=4cm,height=2.5cm]{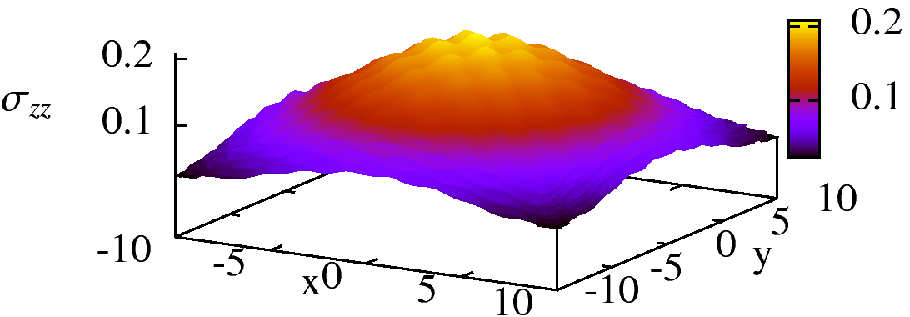}
  \includegraphics[width=4cm,height=2.5cm]{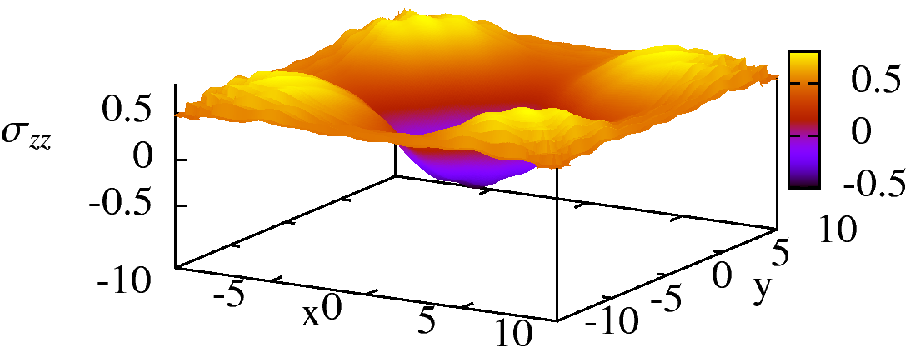}
  \caption{Stress response map, $\sigma_{zz}$, for the bottom of the box of a
    frictional system of size $L=20d$, Left panel: $F_{app}=10mg$. Right
    panel: $F_{app} = 100mg$.}
  \label{fig5}
\end{figure}

\begin{figure}[htbp]
  \includegraphics[width=7cm]{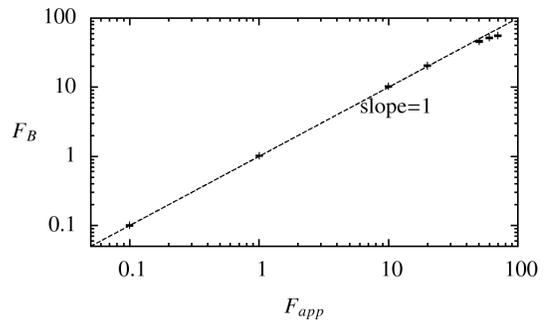}
  \caption{Comparison between the actual applied force $F_{app}$ versus
    $F_{B}$, the fitted force using the Boussinesq equation to fit the stress
    profile data at the bottom of a frictional granular packing of side length
    $L=20d$.}
  \label{fig6}
\end{figure}

The magnitude of the applied force also strongly influences the response
function. For small applied forces, the stress response remains isotropic
consistent with the classical result as seen in the previous figures. Beyond
some force threshold the stress response becomes anisotropic due to the FCC
arrangement of the grains and Eq.~\ref{eq2} is no longer valid as indicated by
the stress maps in Fig.~\ref{fig5}. To quantify this effect and identify the
linear regime as a function of applied force, we performed a similar analysis
as described above. We varied the applied force over several orders of
magnitude, $0.01mg \leq F_{app} \leq 100mg$, applied to a system size of side
length $L=40d$. Figure \ref{fig6} shows that the fitted force increases
linearly with applied force with unit slope when the magnitude of the
perturbation force remains below $F_{app} \simeq 60mg$. In ongoing work to be
reported elsewhere \cite{Lsilbertacakir} we find that this threshold also
depends on the friction coefficient.
Thus, in summary, we find that for ordered frictional packings the stress
state is suitably described by the semi-infinite result even for relatively
small systems over a wide range in applied forcing and we have quantified the
linear regime over which this agreement exists.

\begin{figure}[htbp]
  \includegraphics[width=4cm]{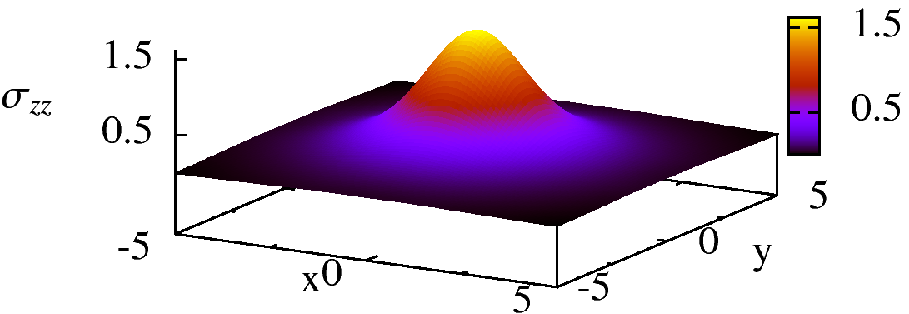}
  \includegraphics[width=4cm]{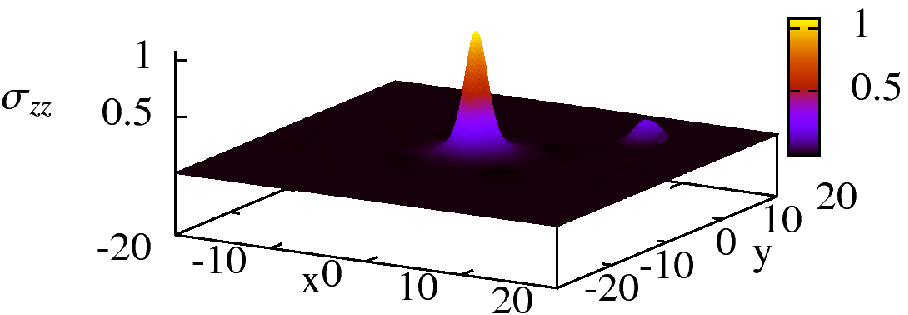}\\
  \includegraphics[width=4cm]{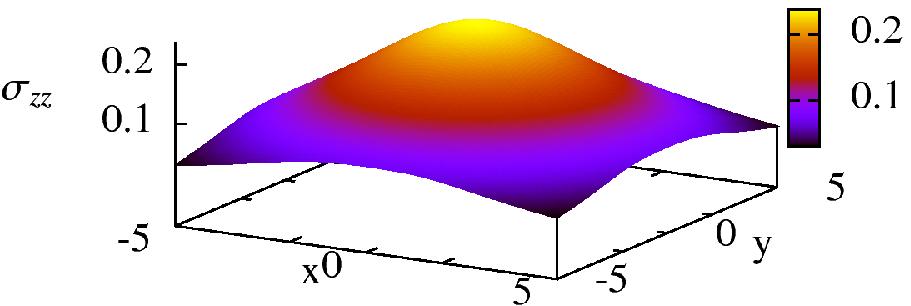}
  \includegraphics[width=4cm]{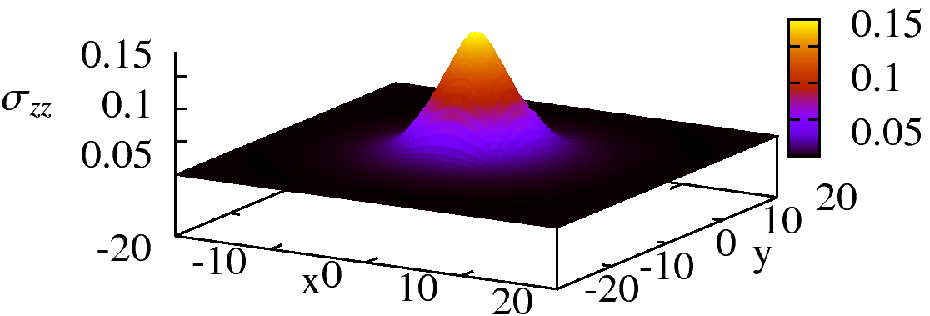}\\
  \includegraphics[width=4cm]{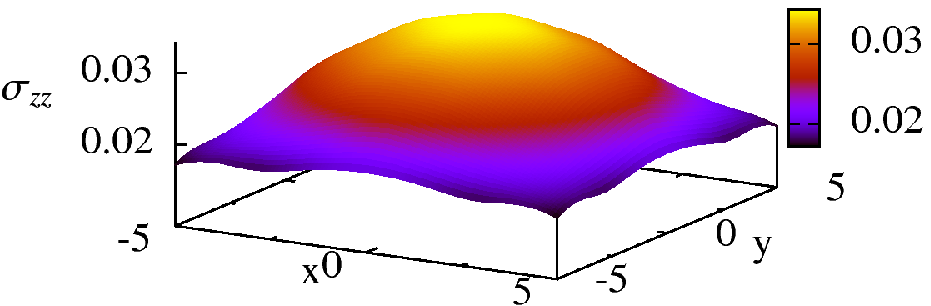}
  \includegraphics[width=4cm]{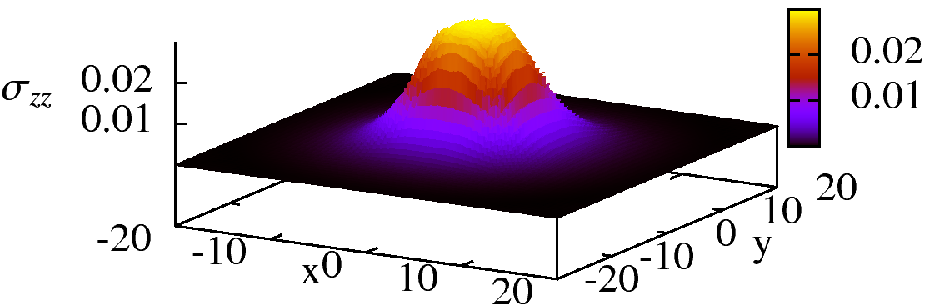}
  \caption{Stress maps, $\sigma_{zz}$, in response to a localized force of
    magnitude $F_{app} = 1mg$ applied at the top of the packing. The panels
    show data at different distances from the perturbation source. Rows show
    data for layer 1 (top), 3, and 8 (near bottom), for $L=10d$ (left column
    panels) and $L=40d$ (right column panels).}
  \label{fig7}
\end{figure}

In an effort to investigate the underlying mechanisms responsible for the
changing character of the stress response in our different systems we have
computed the normal stress response at different distances from the source of
the force perturbation which are shown in Fig.~\ref{fig7}. The left hand
panels in Fig.~\ref{fig7} show data for our smallest system, $L=10d$,
which deviates strongly from the classical predictions despite appearing to
retain a single broad peak response profile. While the right hand columns are
for a larger system, $L=40d$, that is consistent with the Boussinesq
expression. As expected at the top of packing (top panels) the stress is
localized at the point of perturbation. Deeper into the packings the stress
response broadens. For the smallest system the stress quickly spreads across
the entire region of the packing while for larger systems the stress peak
remains relatively localized within the central region of the packing.

\begin{figure}[htbp]
  \includegraphics[width=4cm]{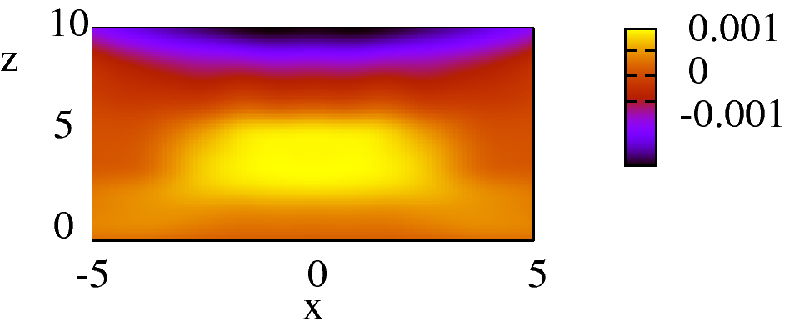}\hspace{0.1 in}
  \includegraphics[width=4cm]{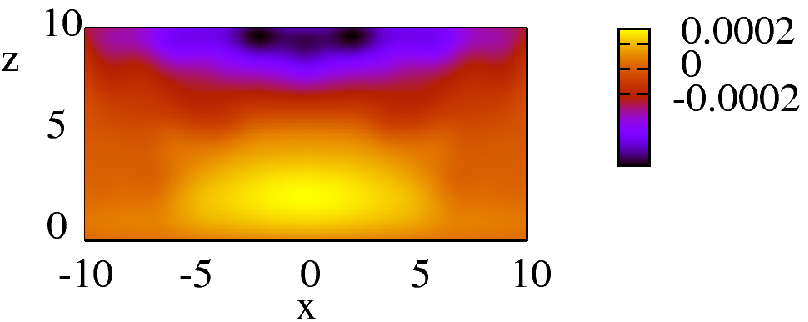}
  \includegraphics[width=4cm]{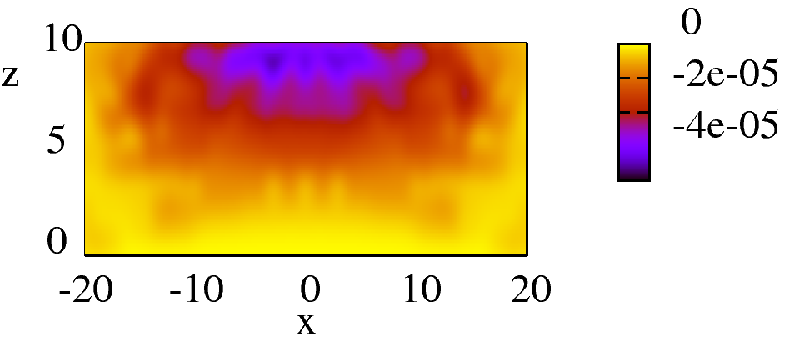}\hspace{0.1 in}
  \includegraphics[width=4cm]{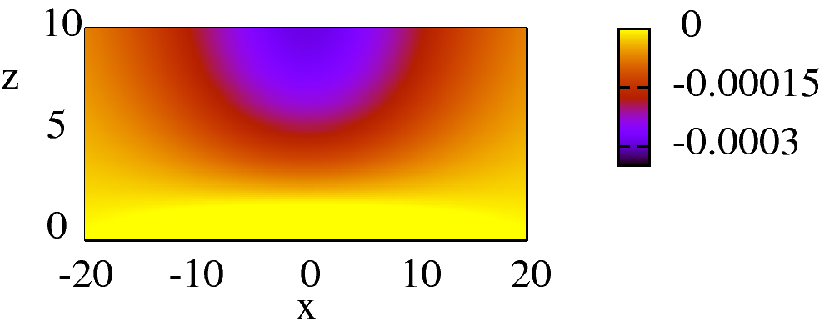}
  \caption{Normal stress, $\sigma_{xx}$, at the side wall for frictional
    packings of height $\approx 10d$ and side length, $L = 10d,~ 20d,~ 40d$
    and the classical expression for a system of size $L=40d$ (bottom right
    panel), using $\nu = 0.4$.}
  \label{fig8}
\end{figure}

The reason behind this changing stress response is a strong reflective
component of stress from the side walls as the stress is transmitted away from
the source of the perturbation. To qualify this effect, in Fig.~\ref{fig8} we
plot the averaged normal stresses, $\frac{1}{2}(\sigma_{xx}+\sigma_{yy}) =
``\sigma_{xx}$'', exerted at the side walls for our different systems, and
data for the classical expression using a value of the Poisson ratio of $\nu =
0.4$. Indeed, we find that for the smaller systems the stress at the side
walls is much larger than that expected from the classical behavior of side
wall stress from Boussinesq equation and the corresponding larger system size.
To generate the data presented in Fig.~\ref{fig8} it is necessary
to introduce the material Poisson ratio $\nu$. Our final choice for $\nu$ was
made on the best fit between the stress profiles obtained via the displacement
fields, Eq.~\ref{eq1}, and the displacement fields constructed from the
simulation data. We matched our simulation displacement fields for our largest
system to the predictions of Eq.~\ref{eq1} using $\nu$ as a fitting
parameter. Our confined packings therefore behave as a continuum material with
an effective Poisson ratio, $\nu \approx 0.4$. This procedure actually
provides a convenient method to determine material properties of composite
materials such as granular packings \cite{PhysRevE.80.061307}.  The benefit of
analyzing the displacement fields is that we can view the direct microscopic
response of the system at the particle scale due to the imposed localized
force perturbation.

\begin{figure}[htbp]
  \includegraphics[width=4cm]{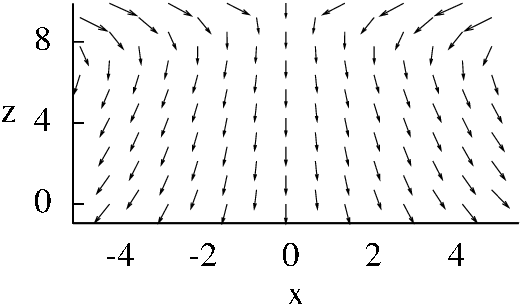}\hspace{0.2 in}
  \includegraphics[width=4cm]{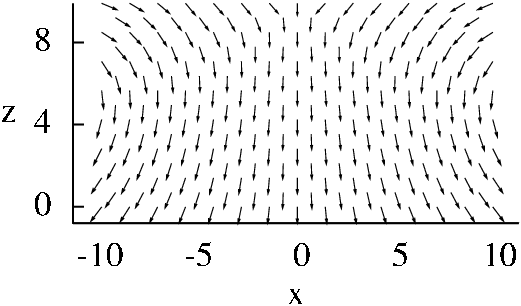}
  
  \vspace{0.1 in}
  \includegraphics[width=4 cm]{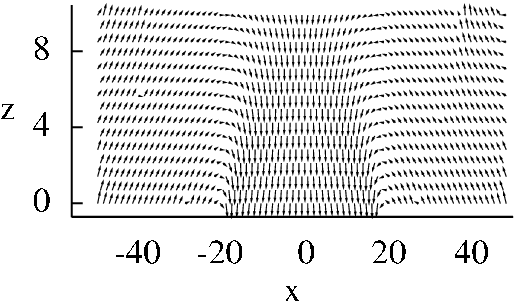} \hspace{0.1 in}
  \includegraphics[width=4 cm]{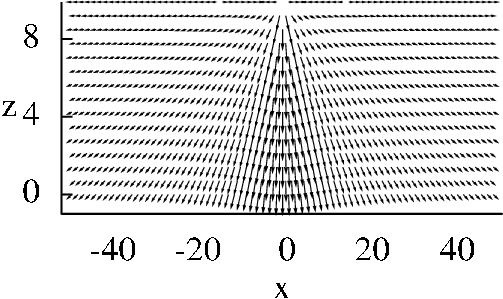}
  \vspace{-0.1 in}
  \caption{Displacement field vectors in the vertical plane for a central
    slice inside packings of side length $L = 10, 20, 100d$, and the result of
    Eq.~\ref{eq1} on a system of size $L=100d$, using $\nu = 0.4$.}
  \label{fig9}
\end{figure}

The displacement fields of the middle vertical slice for frictional packings
of different size are plotted in Fig.~\ref{fig9}. We also include a comparison
to Eq.~\ref{eq1} for a system of size $L=100$, using $\nu = 0.4$. These
changes in the displacement fields of the perturbed system result in a larger
reflection of stress at the side walls hence causing deviations in the overall
stress response of smaller systems compared with the semi-infinite size
result. However, it is worth pointing out that these differences occur only
for relatively small packings of size $L \lesssim 20d $. Above this size,
these frictional packings appear to be suitably described by the Boussinesq
formalism.

\subsection{Frictionless Packings}
\begin{figure}[htbp]
  \includegraphics[width=4cm]{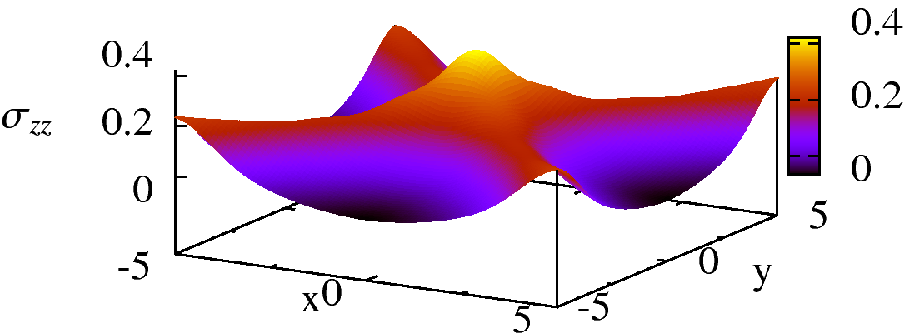}
  \includegraphics[width=4cm]{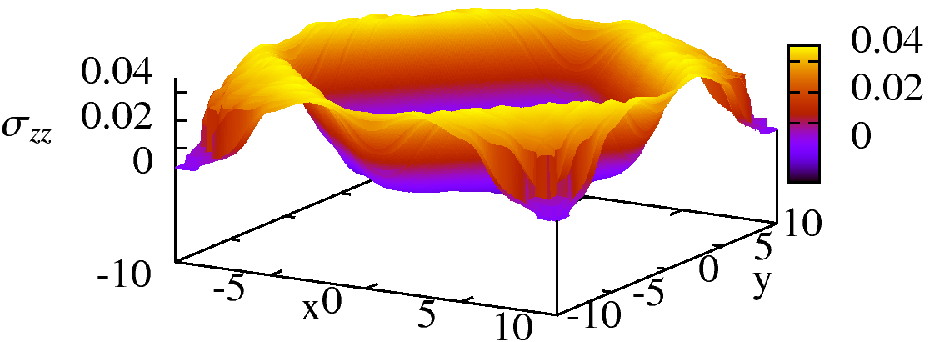}
  \includegraphics[width=4cm]{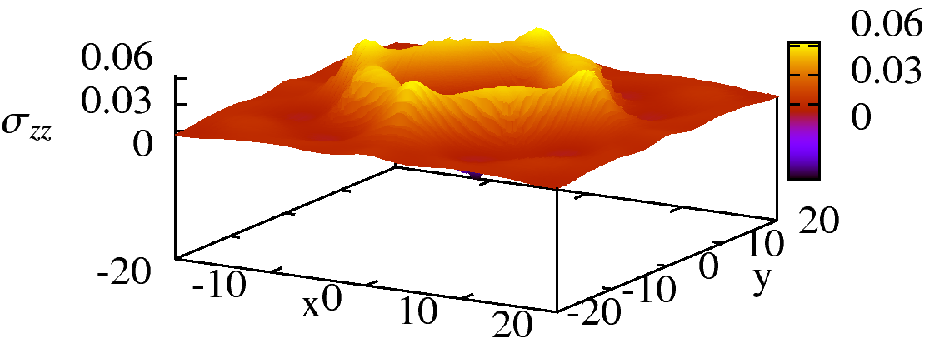}
  \includegraphics[width=4cm]{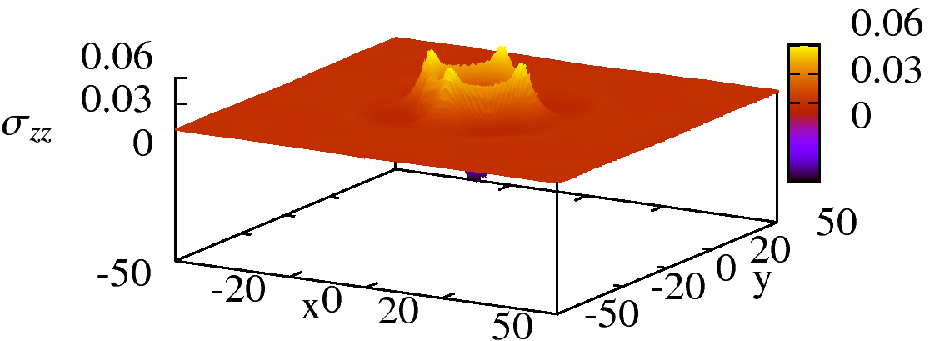}
  \caption{Frictionless systems $\mu = 0$. Stress response maps of
    $\sigma_{zz}$ at the bottom boundary due to an applied force $F_{app} =
    1mg$, for different systems sizes, $10d \leq L \leq 100d$.}
  \label{fig10}
\end{figure}

In contrast to frictional packings, \emph{frictionless} ordered arrays display
primarily strongly anisotropic stress behavior in response to localized force
perturbations. In Fig.~\ref{fig10} the stress component $\sigma_{zz}$ for
frictionless arrays of different sizes are compared. Apart from the smallest
system size, the response profiles are ringed or multi-peaked with a minimum
in the stress response characteristic of an anisotropic response function in
clear contrast to the frictional packings of Fig.~\ref{fig2}. Surprisingly,
however, the smallest system of size $L=10d$ exhibits a stress response that
initially appears isotropic in nature - a maximum in the stress in the middle
of the packing - but with some unusual features not seen in the frictional
case. In fact this smaller system does not conform to the Boussinesq stress
profile even though an isotropic-like single peak response is observed.

\begin{figure}[htbp]
  \includegraphics[width=4cm]{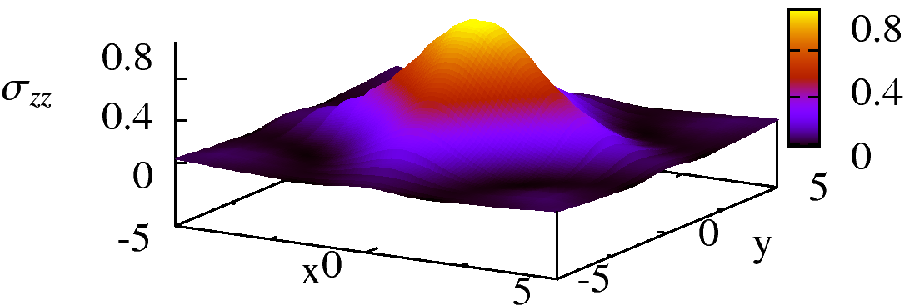}
  \includegraphics[width=4cm]{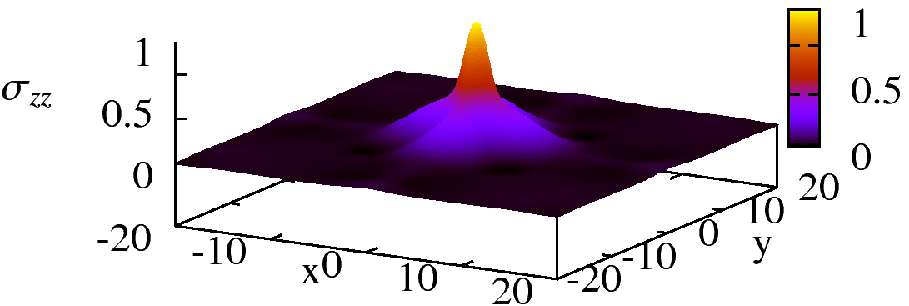}\\
  \includegraphics[width=4cm]{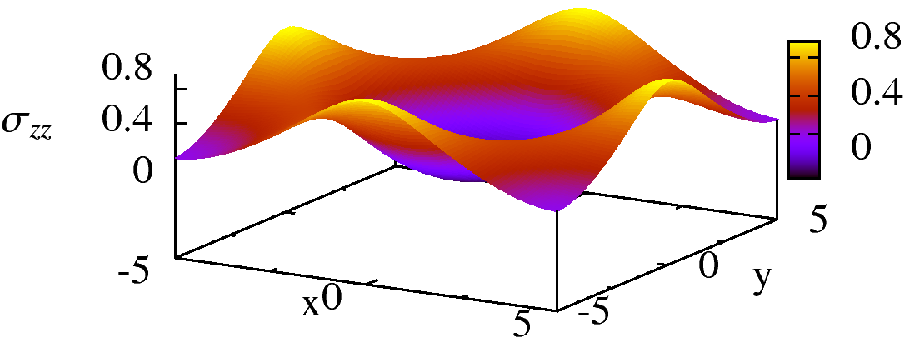}
  \includegraphics[width=4cm]{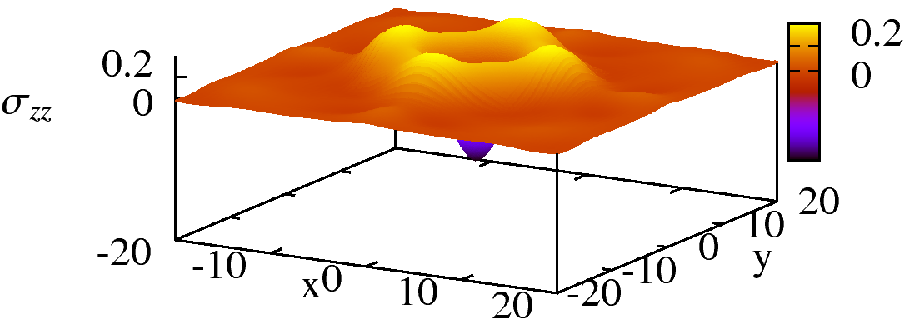}\\
  \includegraphics[width=4cm]{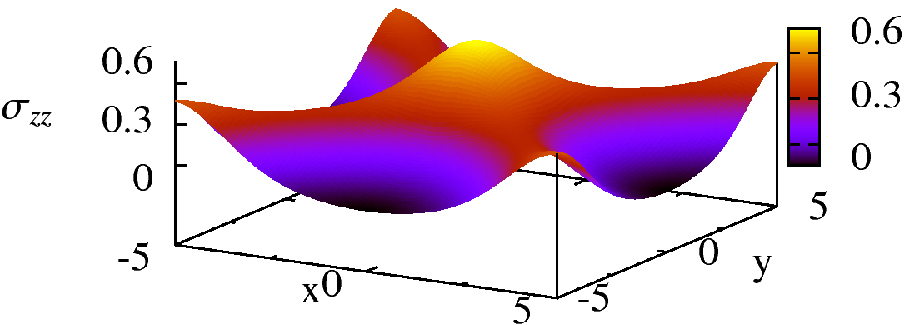}
  \includegraphics[width=4cm]{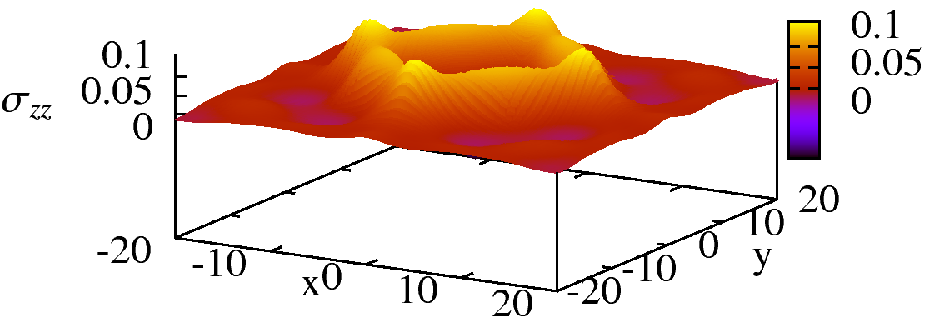}
  \caption{Stress maps, $\sigma_{zz}$, in response to a localized force of
  magnitude $F_{app} = 1mg$ applied at the top of the packing. The panels show
  data at different distances from the perturbation source. Rows show data for
  layer 1 (top), 3, and 8 (near bottom), for $L=10d$ (left column panels) and
  $L=40d$ (right column panels).}
  \label{fig11}
\end{figure}

The influence of confinement plays a significant role in modifying the stress
properties for frictionless packings even within the same packing. To
investigate this effect further, stress response maps at different distances
from the source of the force perturbation are shown in Fig.~\ref{fig11} for
two system sizes, $L = 10d, 40d$. In the vicinity of the localized force
perturbation, the stress response also takes on localized features where the
stress is concentrated in a region directly below the point of application. At
intermediate distances between the top and bottom of the packing both systems
now display strongly anisotropic stress profiles with a stress minimum focused
in the region beneath the point of force application. However, as noted above,
the response measured at the bottom of the packing dramatically changes
character between the two system sizes. These features suggest that the stress
state of small frictionless packings can not only vary substantially within
the packing, but exhibit highly unusual stress properties compared to both
larger frictionless systems and frictional packings.

\begin{figure}[htbp]
  \includegraphics[width=4cm]{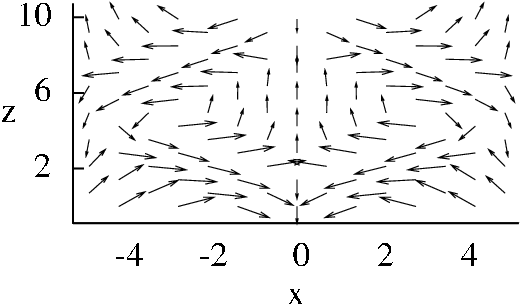}\hspace{0.1 in}
  \includegraphics[width=4cm]{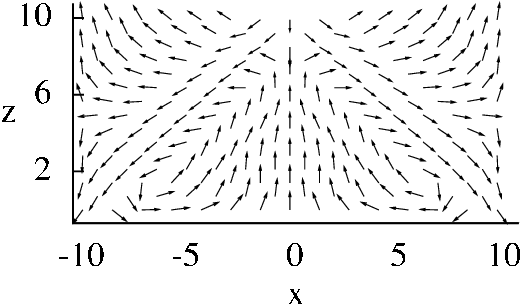}
  \includegraphics[width=4cm]{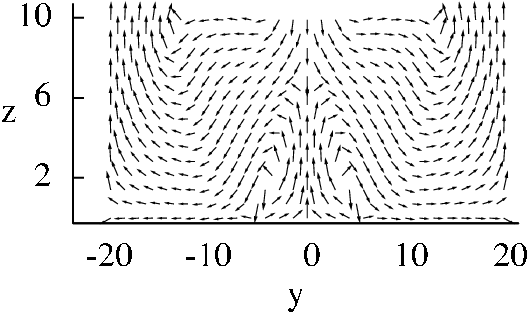}\hspace{0.1 in}
  \includegraphics[width=4cm]{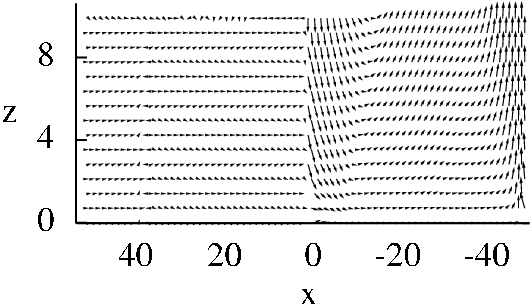}
  \caption{Displacement field vectors in the vertical plane for a central
    slice inside packings of side length $L = 10, 20, 40, 100d$.}
  \label{fig12}
\end{figure}

Again we find that these unusual stress properties are a direct result of the
manner in which stresses are transmitted and interact with the side wall
boundaries due to the strongly correlated displacement fields of the particles
during the response process. Visualization of the displacement fields for
these frictionless packings are particularly illuminating as shown in
Fig.~\ref{fig12}. These displacement vectors are taken from a middle
slice inside the different packings. For the smallest packing (top left panel
in Fig.~\ref{fig12}), the particle displacements in response to the
applied perturbation show a directional structure that is responsible for the
increased stress reflection at the side walls. Interestingly, this reflection
shields the bottom of the packing from the growing stress minimum apparent at
intermediate depths. Whereas, for larger systems the stress response is seen
to be dominated by particle displacements along directed rays through the
system responsible for the anisotropic profiles observed at the bottom of the
packings.

\begin{figure}[htbp]
  \includegraphics[width=4cm]{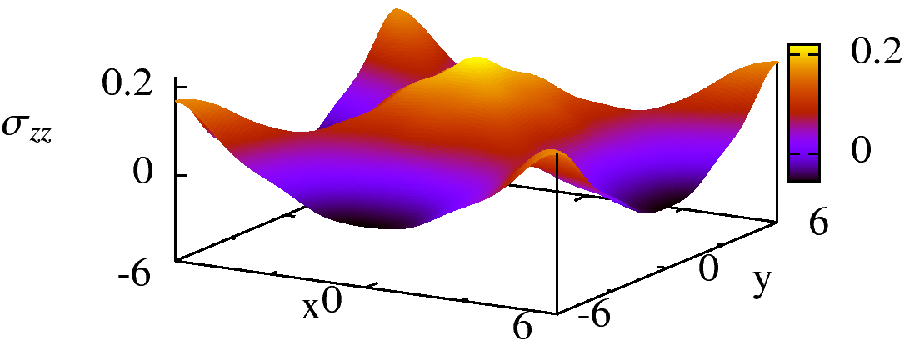}
  \includegraphics[width=4cm]{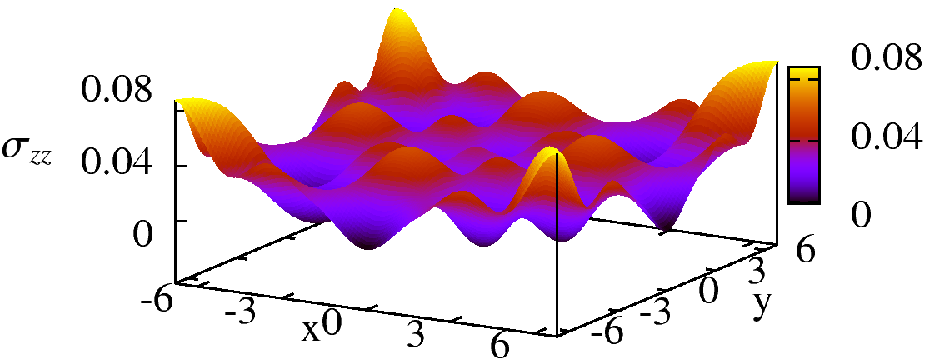}
  \includegraphics[width=4cm]{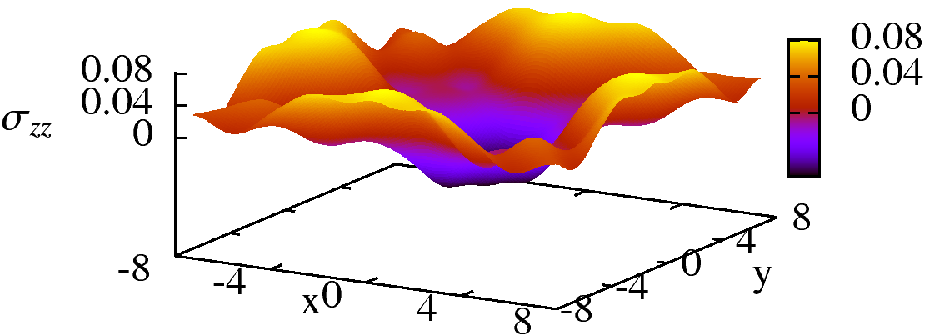}
  \includegraphics[width=4cm]{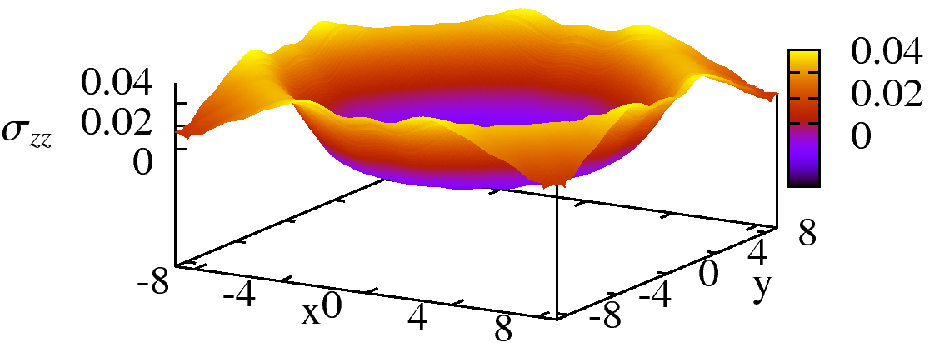}
  \caption{Crossover behavior of the stress response, $\sigma_{zz}$, with
    increasing box size for systems of side length, $12d \leq L \leq 18d$.}
  \label{fig13}
\end{figure}

The results presented above suggest that there exists some crossover response
for confined, frictionless packings that depends on the size of the system. We
have investigated the changing character of the stress response and its
dependence on box size. This crossover behavior is illustrated in stress
response maps of Fig.~\ref{fig13}, where the box side lengths have
been incrementally increased from $L=12d-18d$.
It is apparent that changing box size results in large-scale stress
fluctuations that span the system as the stress response crosses over from a
more isotropic-like response (smaller systems) to strongly anisotropic
behavior (larger system). There is a region of box sizes for these
frictionless packings where the stress state of the system exhibits peculiar
behavior that does not conform to either a pseudo-isotropic description nor
can it be considered anisotropic in character. These features can have
implications in the design of mechanical systems at small scales where
confinement can have a large effect on the stress state of the system.

\begin{figure}[htbp]
  \includegraphics[width=4cm]{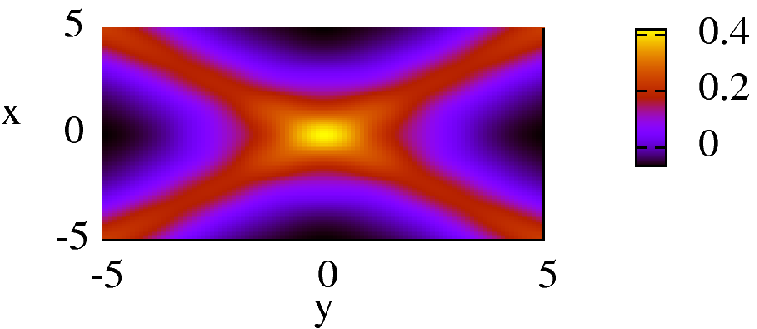}  
  \includegraphics[width=4cm]{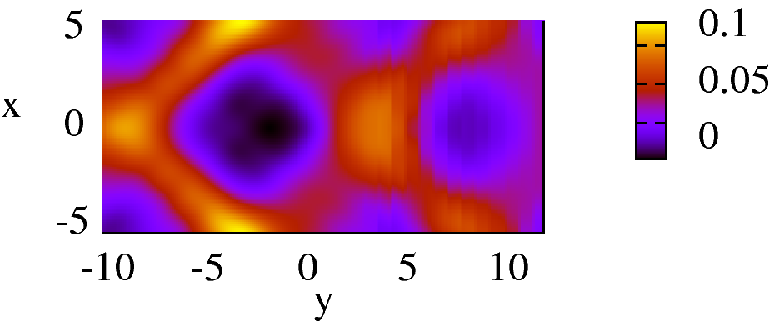}
  \includegraphics[width=4cm]{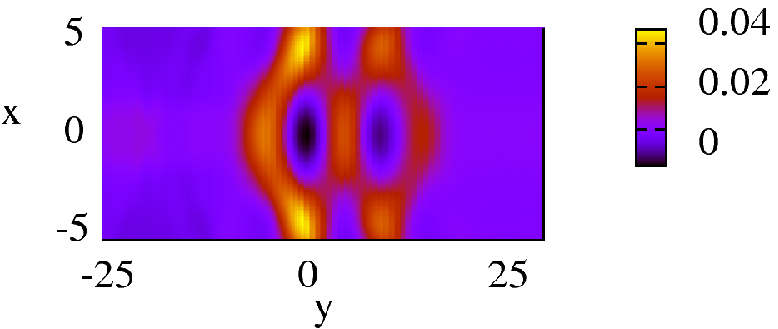}
  \includegraphics[width=4cm]{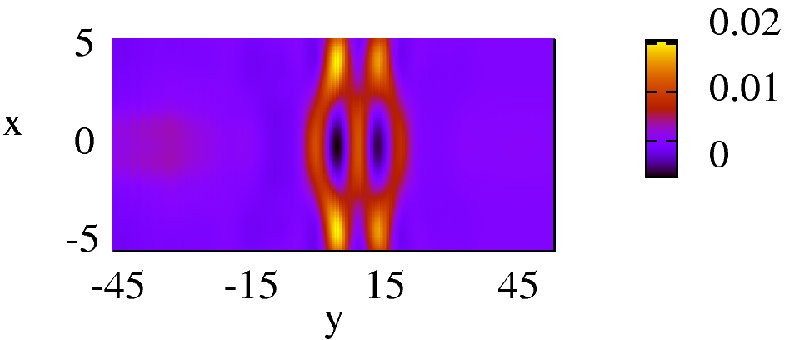}
  \caption{Evolution of the stress response, $\sigma_{zz}$, in frictionless
    packings as one side of the confining box is increased from a square base
    of side lengths $L_{x} = L_{y} = 10d$, to a rectangular base of side
    lengths $L_{x} = 10d$ and $L_{y} = 23d$, $L_{x} = 10d$ and $L_{y} = 50d$,
    $L_{x} = 10d$ and $L_{y} = 100d$.}
  \label{fig14}
\end{figure}

We have also investigated the role of box geometry on the nature of the stress
response. As shown earlier we find that for the smaller frictionless systems
confined within a square-base box with $L_{x} = L_{y} = 10d$, the stress
response exhibits a broad peak in the central region of the base. We find that
highly unusual and complex features emerge if we now increase the size of only
one side of the box into a rectangular base keeping $L_{x} = 10d$
fixed. Stress maps for this scenario are shown in Fig.~\ref{fig14},
where we have increased the side ratio from 1 to 10. These result emphasize
the importance of box size and geometry on the nature of stress response in
shallow packings.

\section{Conclusions}
Using computer simulations of a model granular material, we have studied the
nature of the stress response in shallow and confined granular crystalline
arrays. Packings composed of highly frictional particles generally exhibit a
stress response function measured at the bottom of the packing that resembles
that of an isotropic, elastic material that is well described by the
Boussinesq equations of classical elasticity. Thus, the semi-infinite
half-space result remains suitably valid to describe the stress state of the
system provided the size of the base of the confining walls is larger than
approximately 40 particle diameters. In this regime the stress response
appears to behave as a linear elastic material. For smaller frictional
systems, the stress response exhibits confinement effects that are not
accurately captured by the classical result. Packing arrangement and the
influence of the boundaries have significant effects on the stress state of
the system. Furthermore, due to the FCC crystalline arrangements studied here,
anisotropic stress response behavior is observed at large applied forces
whereby the influence of structure starts to dominate the stress properties.

Frictionless packings exhibit stress response properties that are generally
anisotropic in nature due to the FCC arrays. However, smaller packings can be
tuned to observe a range of behaviors depending on the location within the
packing for square boxes, or on the ratio of the side lengths for rectangular
systems. Complex response patterns emerge for frictionless rectangular
arrays. We also point out that for the anisotropic profiles - either with zero
friction or for large forcing - the resulting stress response indicates a
decrease in the stress relative to the initial state prior to perturbation,
i.e. in the measures defined here the stress becomes negative. This suggests
that such force perturbations actually weaken the material relative to the
unperturbed state.

\begin{figure}[htbp]
  \includegraphics[width=7cm]{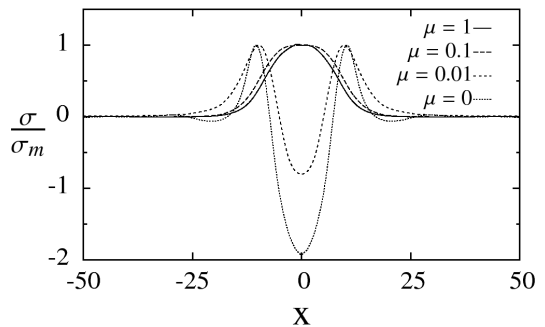}
  \caption{$\sigma$, averaged stress profiles for different friction
    coefficients, $\mu$, for one system size, $L=100 d$, in response to an
    applied force $F_{app} = 1mg$. Each stress profile is scaled by its
    maximum value, $\sigma_{m}$}
  \label{fig15}
\end{figure}

Our results further suggest that there exists a transition in the underlying
character of the response as the friction coefficient is varied. To highlight
this possibility, we show in Fig.~\ref{fig15}, the averaged stress profiles
for a square-base, $L=100d$, for different particle friction coefficients in
response to an applied force, $F_{app} = 1mg$. Indeed we find that increasing
the friction coefficient suppresses the stress dip in the central region of
the packing until a single-peak response is recovered for moderate to large
friction coefficients. We expect that the precise location of this transition
will also depend on the magnitude of the applied force \cite{Lsilbertacakir}.

Our studies indicate that for confined granular arrays, particle properties,
magnitude of forcing, and geometry of the confining container can all be used
to design materials with specific stress states. Through this accessible
parameter space, confined granular packings can be made to mimic the
properties of linear, isotropic, elastic materials, where the stress is
concentrated beneath the point of perturbation - single-peak response. Or they
can be designed to exhibit strongly anisotropic properties where stresses are
transmitted along preferred directions. Moreover, for particular box
geometries, the stress response can be transmitted approximately uniformly
through the system. Thus, it might be necessary to take into account the
possible stress states indicated here during the design of microscopic
granular devices and components.

\section{Appendix}
\label{appendix}
Here we provide a short derivation of the normal stress response component
$\sigma_{zz}$ (see Eq.~\ref{eq2}), within the framework of linear elasticity,
indicating how the material parameters, $G$ and $\nu$, cancel from the
expression given in Eq.~\ref{eq2}. This derivation follows directly from
Landau and Lifshitz \cite{Landau}. The basic ingredient is the implementation
of a constitutive relation between stress and strain that corresponds to
Hooke's law: the normal stress $\sigma_{zz}$ is linearly proportional to the
strain components $u_{xx}$, $u_{yy}$, and $u_{zz}$,
\begin{equation}
  \sigma_{zz}=\frac{2 G}{1-2 \nu}\left[u_{zz}+\nu(u_{xx}+u_{yy}-u_{zz})\right]
  \label{hookeszz}
\end{equation}
where the strain components are obtained from derivatives of the components of
the displacement fields leading to
\begin{equation}
  \begin{tabular}{ccl}
    $u_{xx}$&$=$&$\frac{1}{4\pi G}\left[\frac{z}{\rho^{3}}-\frac{3 x^2 z
      }{\rho^{5}}-\frac{1-2\nu}{\rho^{2}+\rho z}-\frac{(1-2 \nu)x^2(2+\frac{z}{\rho})}{(\rho^{2}+\rho z)^2}\right]$, \\ \\
    $u_{yy}$&$=$&$\frac{1}{4\pi G}\left[\frac{z}{\rho^{3}}-\frac{3 y^2 z
      }{\rho^{5}}-\frac{1-2\nu}{\rho^{2}+ \rho z}-\frac{(1-2 \nu)y^2(2+\frac{z}{\rho})}{(\rho^{2}+\rho z)^2}\right]$, \\ \\
    $u_{zz}$&$=$&$\frac{1}{4\pi G}\left[\frac{2 \nu z}{\rho^{3}}-\frac{3 z^3}{\rho^{5}}\right]$,
  \end{tabular}
  \label{strains}
\end{equation}
and we have set the value of the applied, perturbing force to unity for
notational convenience.

Combining the above equations we find, 
\begin{align}
  \sigma_{zz}&=\frac{2 G}{1-2 \nu}\frac{1}{4\pi G}[\frac{2 \nu
    z}{\rho^{3}}-\frac{3 z^3}{\rho^{5}}+ \nu(\frac{2 z}{\rho^{3}}-\frac{3
    (x^2+y^2) z}{\rho^{5}}-
  \nonumber \\
  &\frac{2 (1-2\nu)}{\rho^{2}+\rho z}-\frac{(1-2 \nu)(x^2+y^2)
    (2+\frac{z}{\rho})}{(\rho^{2}+\rho z)^2}-\frac{2 \nu z}{\rho^{3}}+\frac{3
    z^3}{\rho^{5}}]\nonumber
\end{align}
explicitly emphasizing the cancellation of $G$ from the equation.

The resulting expression can be written as
\begin{align}
  \sigma_{zz}&=\frac{1}{2 \pi (1-2 \nu)}[\frac{\nu z}{\rho^{3}}(4-2 \nu
-3)+\frac{z^{3}}{\rho^{5}}(3 \nu +3\nu -3)-
\nonumber \\
&\frac{2 \nu (1-2 \nu)}{\rho^{2}+\rho z}-
\frac{\nu(1-2\nu)(1-\frac{z}{\rho})(2+\frac{z}{\rho})}{\rho^{2}+ \rho z}]
\nonumber
\label{hookeszzstep1}
\end{align}
and with further algebra,
\begin{equation}
  \sigma_{zz}=\frac{1}{2 \pi}\left[\frac{\nu z}{\rho^{3}}-\frac{3 z^{3}}{\rho^{5}}-\frac{1}{\rho^{2}+r z}(2
    \nu-2 \nu+ \frac{\nu z}{\rho}+\frac{\nu z^2}{\rho^{2}})\right] \nonumber
\label{hookeszzstep2}
\end{equation}
The terms containing $\nu$ cancel out and one obtains Eq.~\ref{eq2} for an
arbitrary applied force, $F_{\rm app}$, directed in the $-z$ direction.

\section{Acknowledgements}
This work is dedicated to Isaac Goldhirsch.  We acknowledge support from the
National Science Foundation CBET-0828359.


\begin{thebibliography}{29}
\bibitem[{nor()}]{normal}
\bibinfo{note}{The coarse graining function is integrated over all space then
  normalized to the value of the applied force $F_{app}$.}
\end{thebibliography}
\end{document}